%% file: main.tex
\PassOptionsToPackage{table,xcdraw}{xcolor}
\documentclass[acmsmall]{acmart}
\usepackage{booktabs}
\usepackage{fontawesome5}
\usepackage{bbding}
\usepackage{pifont}
\usepackage{wasysym}
\usepackage{enumitem}
\usepackage{gensymb}
\usepackage{url}
\usepackage{multirow}
\usepackage{algpseudocode}
\usepackage[product-units=repeat]{siunitx}
\usepackage{graphicx}
\usepackage{makecell}
\newcommand{\revise}[1]{#1}

\AtBeginDocument{%
  \providecommand\BibTeX{{%
    \normalfont B\kern-0.5em{\scshape i\kern-0.25em b}\kern-0.8em\TeX}}}

\setcopyright{cc}
\setcctype{by}
\acmJournal{PACMHCI}
\acmYear{2026} \acmVolume{10} \acmNumber{8} \acmArticle{ISS021}
\acmMonth{11} \acmDOI{10.1145/3859333}

\begin{document}

\title{GraspPressure: Understanding Pressure Input With Busy Hands}

\author{Anand Kumar}
\orcid{0009-0002-6387-0604}
\affiliation{%
  \institution{University of Calgary}
  \streetaddress{2500 University Dr NW}
  \city{Calgary}
  \country{Canada}
  \postcode{T2N 1N4}}
\email{anand.kumar1@ucalgary.ca}

\author{Aditya Shekhar Nittala}
\orcid{0000-0002-3698-9733}
\affiliation{%
  \institution{University of Calgary}
  \streetaddress{2500 University Dr NW}
  \city{Calgary}
  \country{Canada}
  \postcode{T2N 1N4}}
\email{anittala@ucalgary.ca}

%%
% Altering Taste Perception of Beverages with Electrotactile Stimulation
% Understanding the 

\begin{CCSXML}
<ccs2012>
   <concept>
       <concept_id>10003120.10003121.10011748</concept_id>
       <concept_desc>Human-centered computing~Empirical studies in HCI</concept_desc>
       <concept_significance>500</concept_significance>
       </concept>
 </ccs2012>
\end{CCSXML}

\ccsdesc[500]{Human-centered computing~Empirical studies in HCI}

\begin{abstract}

In many real-world scenarios, users interact with digital systems while their hands are partially occupied, such as when holding tools or everyday objects. Prior work on busy-hand interaction has focused on microgestures and motion-based techniques, which rely on finger articulation or relative movement. In situations where a stable grasp is required, such as handling fragile, heavy, or precision objects, such motion can be undesirable or impractical. Pressure-based input offers a complementary modality that enables interaction without changing finger pose, yet its use while holding objects has received limited empirical attention. To explore this, we conducted two within-subject experiments examining users’ ability to perform controlled pressure selections and analyzing finger-specific performance across different grasps and geometries. Results indicate that participants can reliably perform pressure-based selections in these conditions, achieving high success rates (92.7\% on average) with low selection times (1.74s on average). Our findings characterize the potential of pressure input as a complementary microinteraction modality under grasp-constrained, busy-hand scenarios.

\end{abstract}

%\keywords{wearable devices; epidermal devices; survey; soft wearables}

\keywords{Force Input, Pressure Input, User Study, Grasp Types, Busy-hand interactions}

\begin{teaserfigure}
  \includegraphics[width=\textwidth]{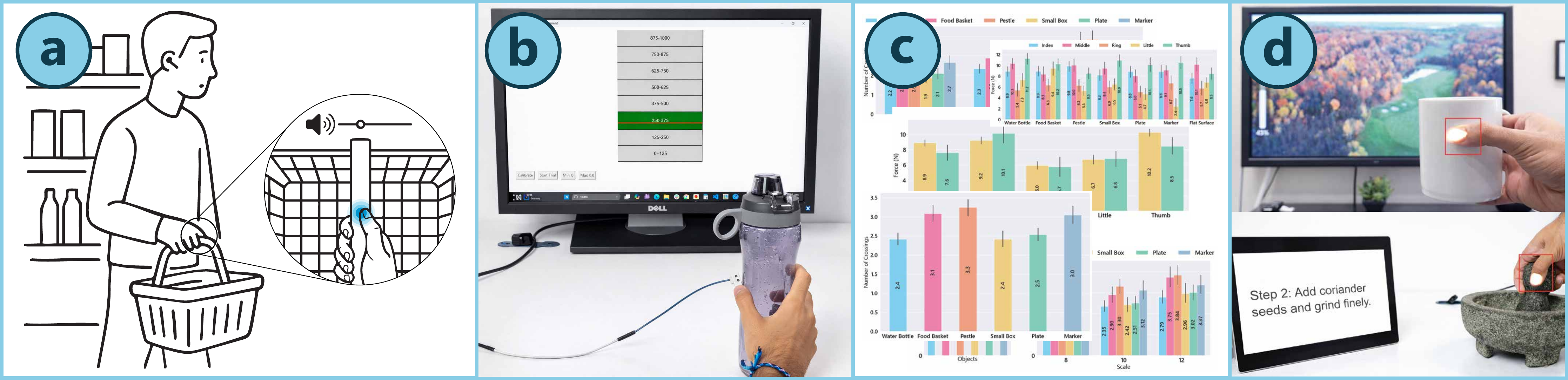}
  \caption{Overview of our exploration of pressure-based input in busy-hand scenarios: (a) motivating example highlighting the potential of pressure sensing while holding objects, (b) experimental setup for capturing pressure input data, (c) analysis of pressure-based input performance across objects and fingers, and (d) application scenarios demonstrating real-world feasibility.}
  \label{fig:teaser}
\end{teaserfigure}

\maketitle

\input{Introduction}
\input{related_work_v2}
\input{experiment1_v3}
\input{experiment2_v2}
\input{applications}
\input{discussion}
\input{conclusion}

\begin{acks}

We thank the anonymous reviewers for their valuable feedback. This work was supported by the Natural Sciences and Engineering Research Council of Canada (NSERC; RGPIN-2023-03608), A-MEDICO (Alberta Medical Devices Innovation Consortium), and the University of Calgary. We also thank Antony Albert Raj Irudayaraj for his contributions to the initial ideation and study process.

\end{acks}

%\section{Working Principle}
%\subsection{Electrochemical Sensing}
%\subsection{Electrotactile Stimulation}

%\balance
\bibliographystyle{ACM-Reference-Format}
\bibliography{sample-base}

% \appendix
% \section{Appendix}

\end{document}

%% file: Introduction.tex
\section{Introduction}
% In many real-world contexts, users frequently interact with digital systems while their hands are partially occupied such as holding tools, carrying objects, or performing manual tasks. Traditional input modalities such as touchscreens or buttons can become impractical in such “busy-hands” scenarios due to limited dexterity, restricted mobility, or the need to maintain continuous contact with the object being held. A common input modality that is actively being investigated for interaction with busy hands is gestural and microgestural interaction~\cite{grasping_microgestures_chi19,sharma2021solofinger,graspui_dis24,wolf2011taxonomy}. These microgestures have been shown to be useful for direct and subtle interactions~\cite{sharma2021solofinger,sparseimu_tochi23}. 

In many real-world scenarios, users interact with digital systems while their hands are occupied, e.g., holding tools, carrying objects, or performing manual tasks. Traditional input modalities such as touchscreens or buttons become impractical due to limited dexterity, restricted mobility, and the need to maintain a stable grasp. A commonly explored solution is gestural and microgestural interaction~\cite{grasping_microgestures_chi19,sharma2021solofinger,graspui_dis24,wolf2011taxonomy}, which enables subtle, direct, and discrete input~\cite{sharma2021solofinger,sparseimu_tochi23}.

While microgestures effectively support discrete interaction with busy hands, pressure input, despite being extensively studied in HCI, remains underexplored in this context. Pressure enables continuous, low-effort interaction while holding objects, without requiring changes in hand posture~\cite{pressure_input-mobile-devices_chi10}. It supports eyes-free control~\cite{pressure_menu_selection_mobilehci10}, fine-grained input~\cite{presstures_chi14,pressure_menu_selection_chiea'24}, and remains effective even when users are mobile or focused on a primary task (Figure~\ref{fig:teaser}(a))~\cite{pressure_walking_mobilehci11}. It is also more resilient to false activations due to its intentional nature, particularly at higher force levels. From an engineering perspective, pressure sensing relies on low-cost, widely available hardware~\cite{micropress_sui22,activeedge_chi19,wristflex_gesture_input,force_picker_chi18}. Based on these benefits, we believe that pressure input can act as a complementary modality and warrants further study in the context of busy-hand interactions.

% However, we still lack a systematic investigation of a more comprehensive set of object geometries and their respective grasps to examine the complex relationship between handheld objects and the pressure input required when holding them and performing a primary task.

However, we still lack a systematic investigation across diverse object geometries and associated grasps to understand how they influence pressure input during concurrent task performance.

% Several empirical investigations in the HCI literature studied how humans perform pressure-based input. These fundamental investigations include examining the ability of users to accurately select absolute force levels~\cite {pressure_input_bimanual_interaction_chi14,pressure_widgets,pressure_menu_selection_mobilehci10,augmenting_mouse_with_pressure_input_chi07,force_controlled_input_chiea05,backxpress_chi17,haptic_feedback_car_autoui16}. However, these majorly involved rigid surfaces such as a mouse~\cite{augmenting_mouse_with_pressure_input_chi07,pressurefish_chi08}, or mobile touchscreens~\cite{characteristics_pressure_input_chi10,backxpress_chi17}. Very few studies have investigated pressure sensing on soft devices. These include understanding pressure input on textile patches~\cite{zpatches_tei18} and tiny wearables~\cite{deformwear_imwut17}. More recent work has investigated the effect of pressure input on soft, malleable and deformable surfaces ~\cite{squish_this_chi21,force_input_deformable_tochi}. However, in all these works, pressure input was performed with non-dominant hands, and none of them examined how holding objects can influence pressure input.

Several empirical studies in HCI have examined how users perform pressure-based input, particularly their ability to accurately select absolute force levels~\cite{pressure_input_bimanual_interaction_chi14,pressure_widgets,pressure_menu_selection_mobilehci10,augmenting_mouse_with_pressure_input_chi07,force_controlled_input_chiea05,backxpress_chi17,haptic_feedback_car_autoui16}. However, most of this work focuses on rigid surfaces, such as a mouse~\cite{augmenting_mouse_with_pressure_input_chi07,pressurefish_chi08} and mobile touchscreens~\cite{pressure_input-mobile-devices_chi10, backxpress_chi17}. Fewer studies have explored pressure input on soft or deformable surfaces, including textile patches~\cite{zpatches_tei18}, wearable devices~\cite{deformwear_imwut17}, and more recently, malleable interfaces~\cite{squish_this_chi21,force_input_deformable_tochi}. However, these studies typically involve non-dominant hand input and do not consider scenarios where users are holding objects. As a result, the impact of grasping and object interaction on pressure input remains underexplored.

This paper addresses the following research questions, which, to the best of our knowledge, have not been explored in prior work that focused on pressure input.

\begin{itemize}
    \item What is the range of force (maximum voluntary force) that users can exert with each of their fingers while their hands are occupied? 
    \item Which fingers are more comfortable for users to perform pressure input?
    \item How well can humans perform pressure input while their hands are occupied, in terms of success rates, interaction times, and sensitivity to different force levels?
\end{itemize}

We present a systematic study of the feasibility of pressure input with busy hands while holding objects across six fundamental grasp types based on Schlesinger's taxonomy~\cite{schlesinger1919mechanische} through two user experiments (Figure~\ref{fig:teaser}(c)). In the first experiment, we compared the amounts of pressure (Maximum Voluntary Force) that each of the fingers can exert on objects. To address our second research question, we collected feedback from participants to determine which fingers were most suitable and preferred. Results show that the thumb achieves the highest MVF and is the most preferred, followed by the middle and index fingers. In the second experiment, we conducted a user study to measure the thumb's performance in pressure input while holding objects. The results of this study show that users can perform pressure input with a success rate of $\sim$92.7\% across all objects. Even with a higher number of levels (12), the success rate remained high (91.31\%). The average selection time $(1.74s)$ falls within the typical range for microgestures, \revise{suggesting its potential suitability for such interactions. These findings provide design guidance for pressure-based input under grasp constraints. \textbf{Taken together, our results provide evidence supporting pressure microinteractions as a promising complementary modality to microgestures for busy-hand input.}}

%Finally, we implemented three application scenarios using state-of-the-art computer vision techniques to demonstrate integration into daily actions and to highlight current limitations of vision systems in reliably detecting pressure input with busy hands (Figure~\ref{fig:teaser}(d)). 

%% file: related_work_v2.tex
\section{Related Work}

\begin{table*}[h!]
  \caption{Comparison of prior pressure-based input systems across devices, techniques, force ranges, and performance, alongside our busy-hand pressure-input results.}
  \centering
  \fontsize{7}{11}\selectfont
\begin{tabular}{p{0.16\textwidth} p{0.22\textwidth} >{\centering\arraybackslash}p{0.04\textwidth} >{\centering\arraybackslash}p{0.06\textwidth} >{\centering\arraybackslash}p{0.08\textwidth} >{\centering\arraybackslash}p{0.04\textwidth} >{\centering\arraybackslash}p{0.07\textwidth} >{\centering\arraybackslash}p{0.065\textwidth}  >{\centering\arraybackslash}p{0.06\textwidth}  }
\textbf{Study} & \textbf{Context} & \textbf{Grasps} & \textbf{Technique} & \textbf{Force Range (N)} & \textbf{Levels} & \textbf{Success Rate (\%)} & \textbf{Selection Time (s)} & \textbf{Crossings} \\
 \midrule
 Squish This~\cite{squish_this_chi21}     & Soft silicone surfaces   &  0  &  QR &  0-5.51           &   20     &    $\sim$94.7    &    $\sim$2.86            &   $\sim$2.58         \\

PressureWidgets~\cite{pressure_widgets}  & Pressure-sensitive stylus   &  1  &  QR &  NR           &   12     &    $\sim$89.0   &     $\sim$0.80          &  $\sim$0.95        \\

BackXPress~\cite{backxpress_chi17}     & Back of device mobile input  &  1  &  Tap &  0-4           &   7     &    69.4   &     3.50          &   NR       \\

Wilson et al.~\cite{pressure_menu_selection_mobilehci10}     & Mobile Pressure input   &  1  &  QR &  0-3.5  &   10     &    85.0   &     $\sim$2.45           &   4.7        \\

PressureFish~\cite{pressurefish_chi08}     & Pressure input on mouse   &  1  &  Dwell &  NR     &   16     &    $\sim$72.0   &     $\sim$1.65           &   $\sim$0.45        \\

Mizobuchi et al.~\cite{force_controlled_input_chiea05}     & Force controlled pen input   & 1 &  Dwell &  0-4     &   10     &    $\sim$89.0   &     $\sim$4.80           &   NR       \\

Cechanowicz et al.~\cite{augmenting_mouse_with_pressure_input_chi07}     & Dual-Pressure input on mouse   &  1  &  QR & NR          &   12     &    58.0   &     $\sim$3.30           &   $\sim$2.6        \\

McLachlan et al.~\cite{pressure_input_bimanual_interaction_chi14} & Input with Non-dominant hand   &  1  &  QR & NR          &   10     &    89.3   &     2.89         &   NR        \\

DeformWear~\cite{deformwear_imwut17} & Tiny wearables & 1 &  Dwell & 0-5  & 3 & 99.7 & $\sim$2.2 &  NR \\ 

\midrule

\textbf{This Work}     &  \textbf{Busy hands holding objects}   &  \textbf{6}  &  \textbf{QR} & \textbf{$\sim$0-15.31} &   \textbf{12}     &    \textbf{$\sim$92.7}    &     \textbf{$\sim$1.74}          &   \textbf{$\sim$2.82}         \\
 
 \bottomrule

\end{tabular}
\end{table*}

\subsection{Pressure Input Performance on Rigid and Soft Surfaces}

Pressure has been widely studied as an input modality across various HCI form factors. Prior work has primarily examined users’ ability to accurately select absolute force levels~\cite{augmenting_mouse_with_pressure_input_chi07,backxpress_chi17,pressure_input_bimanual_interaction_chi14,force_controlled_input_chiea05,pressure_widgets,pressure_menu_selection_mobilehci10,pressurefish_chi08,pressure_input-mobile-devices_chi10, haptic_feedback_car_autoui16}. A typical task involves controlling a cursor via applied force to reach a visual target, with task difficulty manipulated through target size and position, and performance measured using accuracy and completion time. While early work focused on rigid surfaces in mobile contexts (e.g., touchscreens~\cite{backxpress_chi17}, stylus input~\cite{pressure_widgets}), more recent research has explored pressure input on soft and deformable surfaces, such as textile patches~\cite{zpatches_tei18} and silicone-based wearables~\cite{deformwear_imwut17}. Surface compliance can influence users’ perception of applied force~\cite{srinivasan1995tactual}, and has been shown to affect input performance~\cite{squish_this_chi21}. However, despite this body of work, pressure-based input has not yet been empirically studied in busy-hand scenarios.

\subsection{Interaction with Busy Hands}
% The vast majority of work on interaction techniques and pressure sensing assumes that at least one of the user's hands is free for interaction. A relatively small amount of work has examined settings in which the user's hands and fingers are engaged in another task~\cite{no-handed-interaction_gi17}. Fitzmaurice et al. contributed seminal work with the concept of Graspable User Interfaces~\cite{bricke_chi95}. Graspables~\cite{graspables_chi09} presented grasp-recognition as an input medium. Wimmer presented the GRASP model to show how a grasp conveys meaningful contextual information~\cite{grasp_sensing_tei11}. Wolf et al. assessed the ergonomic factors and suitability of microinteractions in an expert study involving three objects (steering wheel, pen, and cash card)~\cite{wolf2011taxonomy,wolf2016microgestures}. Other work has identified gestures focusing on specific objects and geometries, such as steering wheels~\cite{gesturing_steering_automotiveui14}, bike handles~\cite{gestural_interaction_chi11}, textile-based cords~\cite{etextile_microinteractions_chi20}, and pens~\cite{pen-holding-iss19}. Clarke et al. utilized a motion-matching and pointing technique as input while holding an object~\cite{tracematch_ubicomp16,matchpoint_uist17}.

The majority of work on interaction techniques and pressure sensing assumes that at least one hand is free. Relatively little work has examined scenarios where users’ hands are occupied with another task~\cite{no-handed-interaction_gi17,bricke_chi95,graspables_chi09}. Wimmer’s GRASP model further demonstrated how grasp configurations convey contextual information~\cite{grasp_sensing_tei11}. Prior studies have also explored the ergonomics and feasibility of microinteractions across specific objects, such as steering wheels, pens, and cards~\cite{wolf2011taxonomy,wolf2016microgestures}, as well as object-specific interaction techniques for domains including automotive controls~\cite{gesturing_steering_automotiveui14}, bike handles~\cite{gestural_interaction_chi11}, textile cords~\cite{etextile_microinteractions_chi20}, and pens~\cite{pen-holding-iss19}. Clarke et al. further explored interaction while holding objects through motion-matching and pointing techniques~\cite{tracematch_ubicomp16,matchpoint_uist17}. 

%Graspables~\cite{graspables_chi09} presented grasp-recognition as an input medium. 

%Fitzmaurice et al. contributed seminal work with the concept of Graspable User Interfaces~\cite{bricke_chi95}. 

%Heo et al. presented a taxonomy of such constrained scenarios, which illustrated that each type of constraint affords a different set of gestures~\cite{no-handed-interaction_gi17}. Previous work investigated how interactions should be designed when the hands are occupied. 

\subsubsection{Microgestural Interaction with Busy Hands} Microgestures (or micro-interactions) refer to subtle, rapid finger movements that can be performed with minimal disruption to ongoing tasks~\cite{ashbrook2010enabling}. Building on this, Sharma et al. introduced grasping microgestures, enabling direct and subtle interaction while holding everyday objects~\cite{grasping_microgestures_chi19}. Subsequent work has explored the usability of grasping microgestures and the use of everyday objects as input media~\cite{wolf2011taxonomy,aro_mobilehci21,instantUI_its13,user-defined-gestures_dis23,sharma2021solofinger}. Transferable Microgestures further extend this idea by leveraging the Middle, Ring, and Little fingers to support consistent interaction across different grasps and postures~\cite{transferable_microgesturs_uist23}. In social contexts, everyday objects have also been shown to enable subtle and unobtrusive interaction~\cite{subtelety_chi15,glance_mug_tei18,hidden_interactions_chi21,graspui_dis24}. While interaction with busy hands has been explored, prior work has primarily focused on gestural input. Other modalities remain underexplored, motivating the need for a broader investigation of input techniques in these contexts.

\subsection{Sensing Microinteractions with Busy Hands}

To enable input sensing on objects, prior work has explored a range of sensing modalities. Camera-based approaches are widely used~\cite{object_detection_ieee2023}, but often suffer from occlusion during object manipulation~\cite{mueller2018ganerated}. Another line of work leverages object instrumentation, including conductive sensing by attaching objects to sensor boards~\cite{electrick_chi17,touche_ubicomp12} or embedding touch sensors directly onto surfaces~\cite{nittala2020physioskin,print_a_sketch_chi22,objectskin_imwut2017,midas_uist12,protospray_chi20} or body~\cite{tiptext_uist19,intellilining_chi25,augmenting_on_body_touch_chi23}. Researchers have also explored user instrumentation. These include forearm electromyography (EMG) to detect interactions such as pinching while holding objects~\cite{saponas2010making,sparseemg_uist25,nittala2021computational}, capacitive wristbands to recognize force and grasp~\cite{hand_interaction_tei22}, and bio-acoustic sensing approaches~\cite{vibaware_sui23} that classify tap and swipe gestures using learned signal features. Other approaches include IMU-based sensing~\cite{sparseimu_tochi23}, and acoustic wristband systems~\cite{grab_n_go_imwut25}. While these systems effectively sense microinteractions with busy hands, they primarily focus on discrete inputs such as taps, swipes, and microgestures. To our knowledge, pressure-based sensing and interaction across diverse grasp configurations remain underexplored.

\subsection{Studies on Finger Force Production}

There is extensive literature in ergonomics, human factors, and industrial engineering examining the forces that fingers can produce or withstand~\cite{li2025contributions,cort2011maximum}. Prior work has measured maximal voluntary forces in tasks such as finger pulling~\cite{cort2011maximum}, as well as individual finger contributions in pinching and gripping~\cite{li2025contributions,kong2011individual,RADWIN01031992}. Other studies have investigated the role of tactile feedback in force production~\cite{shim2012tactile,jones2006contribution,hager1996nondigital}, and analyzed finger forces during grasping across varying object properties such as cylinder diameter~\cite{amis1987variation}. Relationships between grip force, normal force, contact area, handle geometry, and hand characteristics have also been explored~\cite{seo2008investigation,alam2025relation}. For a comprehensive review, we refer readers to Alam et al.~\cite{alam2025relation}. \revise{Despite this rich body of work, several gaps remain from an HCI perspective. Existing research provides limited insight into how finger force capabilities and preferences relate to pressure-based input while holding objects in different grasp configurations. In particular, the role of grasp configuration has not been systematically examined across diverse everyday grasps and objects. Moreover, although prior work has established differences in force contribution across fingers~\cite{RADWIN01031992}, it does not address how these differences may inform input-finger selection or pressure-level granularity in interactive systems. These gaps motivate a systematic examination of pressure-based input under different grasp constraints.}

% Despite this rich body of work, several limitations remain from an HCI perspective. First, studies are typically restricted to a single posture or grasp, rather than systematically comparing across diverse grasps. Second, they do not measure forces across all fingers in varied object-based interactions. Third, the forces studied (e.g., pull or grip forces) differ fundamentally from the type of localized pressure input considered in interactive systems. 

%% file: experiment1_v3.tex
\section{Exploratory Study: Finger-Specific Pressure Capabilities and User Preferences Across Grasps}

Before evaluating traditional performance metrics for pressure input (e.g., accuracy, speed, and crossings), we first assess its feasibility under different grasp constraints. Prior work has primarily studied single-digit pressure on surfaces~\cite{pressure_widgets,squish_this_chi21,force_input_deformable_tochi} and has also explored pressure input using the thumb and multiple digits~\cite{pressure-walking-feng, Nittala_multi-touch_skin,SkinMarks} in mobile~\cite{lightlink_uist25} or handheld contexts~\cite{multi-digit}, which may not translate to busy-hand scenarios where grasp stability and finger availability vary. We therefore conduct an exploratory study to understand finger-specific capabilities and preferences across common grasps.

To explore the feasibility of pressure input in busy-hand scenarios, we investigate the following research questions:

\begin{itemize}
   \item \textit{What are the ranges of force that the fingers can apply while holding objects, and how do they differ from the baseline condition (i.e. free hand)?}
    \item \textit{Which fingers are preferred by the users for pressure input with busy hands?}
\end{itemize}
\begin{figure*}[!h]
    \centering
    \includegraphics[width=1\linewidth]{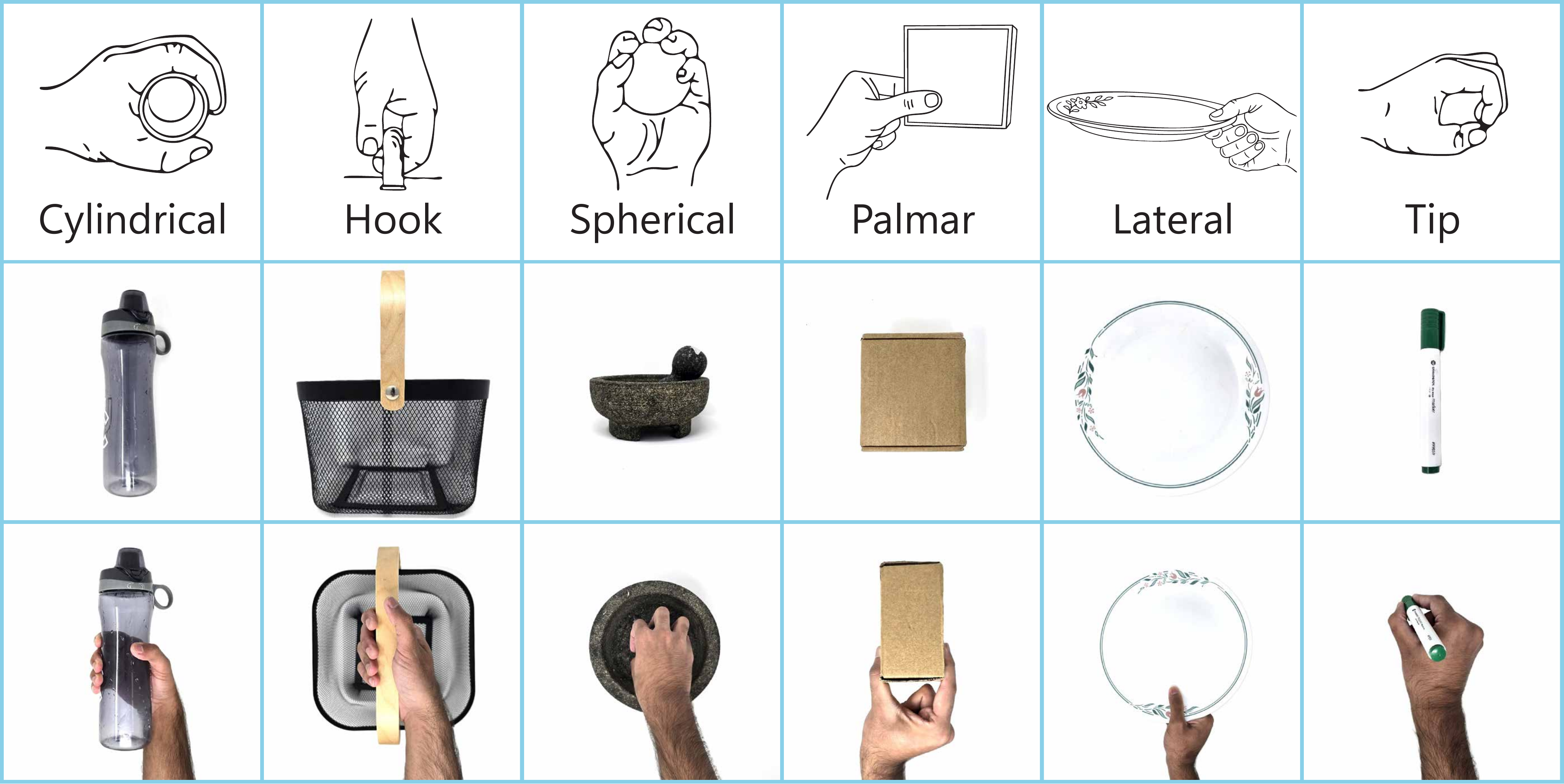}
    \caption{Handheld objects used in the study, selected to represent different grasp types from Schlesinger's grasp taxonomy. Each object corresponds to a primary grasp configuration and was chosen based on prior work on grasp-based interaction~\cite{grasping_microgestures_chi19,heumer2008grasp,sharma2021solofinger}.}
    \label{fig:grasp_types}
\end{figure*}

The first research question examines which fingers provide greater ergonomic suitability for pressure input across different grasp conditions. Based on prior work and common usage patterns, we expect the thumb and index finger to offer higher comfort and feasibility across many grasps. The second question enables a quantitative comparison of force ranges across fingers and grasps, clarifying how object holding and busy-hand constraints affect achievable force and how these ranges vary across digits. From a design perspective, larger force ranges may provide greater tolerance when discretizing pressure into multiple levels, supporting more robust interaction under noisy or constrained conditions. \revise{While extensive biomechanics and ergonomics literature examines finger force production across grasp types, our study differs in two important ways. (1) prior work has focused primarily on the role of individual fingers and their force contributions during grasping, lifting, and object stabilization tasks, often under controlled finger placements and hand postures~\cite{yu2010limits,kong2005optimal}. (2) the task considered here is fundamentally different from conventional grip-force studies. Rather than measuring forces used to maintain a grasp, we measure the maximum voluntary force that an individual finger can exert against an external sensing surface while simultaneously stabilizing an object. This additional interaction constraint introduces a distinct motor context that is directly relevant to pressure-based input. Consequently, while we expect our findings to be broadly consistent with established anatomical and biomechanical trends, we cannot directly infer absolute force thresholds and usable pressure ranges from prior grip-force literature. Our objective is therefore not to establish new physiological principles of finger force production, but to characterize finger-specific pressure-input capabilities under realistic grasp-constrained, busy-hand interaction scenarios.}

%\revise{This exploratory study is not intended to establish novel physiological or biomechanical findings. Rather, it characterizes finger-specific force capability and user preference under the grasp configurations relevant to pressure-based interaction, providing interaction-level guidance for selecting candidate input fingers.}

% This exploratory study is not intended to establish novel physiological findings but to serve as a calibration-oriented investigation that informs the design of pressure-based interactions across grasp types.

\subsection{Grasps}
We based our grasp conditions on Schlesinger’s classification of six prehensile postures, which account for variations in object shape, hand surfaces, and hand configuration~\cite{schlesinger1919mechanische,mackenzie1994grasping}. This taxonomy is widely used in prior work~\cite{grasping_microgestures_chi19,heumer2008grasp,iberall1997human}. The six grasps are:
\begin{itemize}
\item \textit{Cylindrical}: holding cylindrical objects (e.g., a water bottle).
\item \textit{Hook}: grasping heavy loads (e.g., a bag).
\item \textit{Spherical}: holding rounded objects (e.g., a ball or pestle).
\item \textit{Palmar}: grasping with the palm against the object (e.g., a book).
\item \textit{Lateral}: grasping flat objects (e.g., paper, card, or plate).
\item \textit{Tip}: grasping small objects (e.g., a pen or stylus).
\end{itemize}

\subsection{Objects} \label{sec:objects}

We selected six handheld objects corresponding to our six grasp conditions. To ensure coverage across varied everyday contexts, we iteratively compiled a set of representative objects based on prior literature~\cite{wolf2011taxonomy, grasping_microgestures_chi19, liu_grasp_taxonomy_ras2014, pratap2025synergistic} and common daily use. We chose real-world objects over abstract props to help participants relate to realistic pressure-input scenarios~\cite{grasping_microgestures_chi19}. All objects were non-deformable to ensure consistent pressure sensor response. The final set (Figure~\ref{fig:grasp_types}) includes: \textit{a water bottle (cylindrical), a food basket (hook), a pestle (spherical), a filled small cardboard box (palmar), a plate (lateral), and a marker (tip)}.

\noindent
\begin{minipage}[t]{0.75\textwidth}
  \vspace{0pt} % Aligns the top of the text minipage with the top of the image
  \subsection{Sensing and Calibration}
  We used the Interlink FSR400\footnotemark, a commercial force sensor smaller than a fingertip (5.08 mm diameter), to match prior work~\cite{augmenting_mouse_with_pressure_input_chi07, pressurefish_chi08, squish_this_chi21, pressure_input-mobile-devices_chi10} and support replicability~\cite{transparency_CHI_research_chi20}. Participants applied normal force directly to the sensor using a single finger. Following prior designs~\cite{squish_this_chi21,pressure_input_bimanual_interaction_chi14,pressure_input-mobile-devices_chi10}, we implemented a transimpedance circuit using an LM324N op-amp and an Arduino Uno for data acquisition. The Arduino was connected via USB (9600 baud) to a Dell XPS desktop (Intel Core i7, 2.20 GHz), where Python-based software logged sensor data.

  \subsubsection{Calibration}
    Due to the non-linear response of FSRs, we calibrated the sensor to map output values to force (N) following prior work~\cite{squish_this_chi21}. We recorded sensor readings under known weights and fit smooth non-linear functions to obtain calibration curves, enabling a linearized mapping from sensor output to applied force.

    % \subsection{Method}
    
\end{minipage}% <--- Note the percent sign here to prevent a line break
\hfill
\begin{minipage}[t]{0.22\textwidth}
  \vspace{0pt} % Aligns the top of the image minipage with the top of the text
  \centering
  \includegraphics[width=\textwidth]{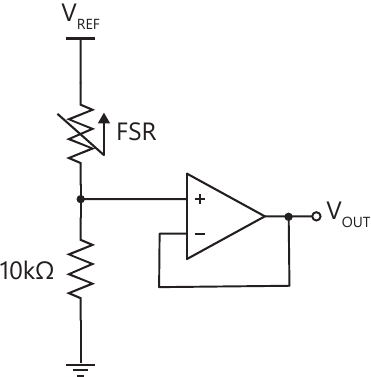}
  \captionof{figure}{Schematic of the force-sensitive resistor (FSR) circuit used to capture the applied force during the experimental tasks.}
  \label{fig:fsr_circuit}
  \Description{Schematic diagram showing the wiring and electrical layout of the force-sensitive resistor circuit.}
\end{minipage}
\footnotetext{Interlink. 2019. FSR 400. \url{https://www.interlinkelectronics.com/fsr-400}}

\subsection{Method}

We recruited 12 participants (8 male, 4 female; mean age = 26, SD = 4.92). Participation was voluntary. The study followed a within-subject design with seven conditions: six objects and a baseline (empty hand). Object order was counterbalanced. For each object, participants held the object naturally with their dominant hand. We then placed five FSRs at the most accessible finger-specific locations based on each participant’s grasp. To account for variability in hand size and grasp strategy, sensor placement was not fixed across participants. If a finger could not comfortably apply force, it was excluded. This ensured participants could exert maximum force without constraints from suboptimal placement.

Participants first indicated their preferred finger for pressure input. They then rated the comfort of each finger on a 7-point Likert scale (1 = least comfortable, 7 = most comfortable), following prior work on micro-interactions~\cite{digitspace_huang_chi16}. Next, they applied maximum force with each finger, referred to as Maximum Voluntary Force (MVF). Maximum Voluntary Force is the highest force a muscle or group of muscles can produce through conscious effort~\cite{shim2012tactile}. Finger order was randomized for each object. In the baseline condition, participants exerted MVF with each finger on a table. The study comprised 7 (6 Objects + Baseline) $\times$ 5 (Fingers) $\times$ 4 (Trials for Maximum Force) $\times$ 12 (Participants) = 1680 trials. Sessions lasted 30–40 minutes \revise{and the study was reviewed and approved by our institutional ethics board. All participants provided informed consent before participation and were informed that participation was voluntary and that they could withdraw at any time. Participants were given breaks between trials, and the experiment was stopped if a participant experienced discomfort.}

\subsection{Results}

Because our data were not normally distributed, we applied the Aligned Rank Transform (ART) of Wobbrock et al.~\cite{wobbrock2011aligned}. For each factor of interest, we aligned and ranked the data with the ART tool~\cite{wobbrock2011aligned} and analyzed them with a repeated-measures ANOVA; we applied the Greenhouse-Geisser correction when sphericity was violated. We followed significant effects with pairwise post-hoc comparisons (paired $t$-tests on per-participant means with Holm correction) at a 95\% confidence level. We analyzed comfort ratings with a Friedman test followed by Nemenyi post-hoc comparisons.

\noindent
\begin{minipage}[t]{0.47\textwidth}
  \vspace{0pt} % Aligns the top of the text minipage with the top of the image

    \subsubsection{\textbf{Force Thresholds}}
    Figure~\ref{fig:force_thresholds_finger} shows average force thresholds across fingers, objects, and participants. Mean force values were: \textsc{Thumb}: 9.96N (SD = 4.14N), \textsc{Index}: 8.71N (SD = 3.58N), \textsc{Middle}: 9.20N (SD = 4.31N), \textsc{Ring}: 5.62N (SD = 4.28N), \textsc{Little}: 5.43N (SD = 4.54N). A repeated-measures ANOVA on the aligned-ranked data revealed a significant main effect of \textsc{Finger} on \textsc{Force Thresholds} ($F_{4,44} = 9.28, p < 0.0001$). Post-hoc comparisons (paired $t$-tests, Holm-corrected) showed that the fingers separated into a high-force cluster (\textsc{Thumb}, \textsc{Index}, \textsc{Middle}) and a low-force

\end{minipage}% <--- Note the percent sign here to prevent a line break
\hfill
\begin{minipage}[t]{0.5\textwidth}
  \vspace{0pt} % Aligns the top of the image minipage with the top of the text
  \centering
  \includegraphics[width=\textwidth]{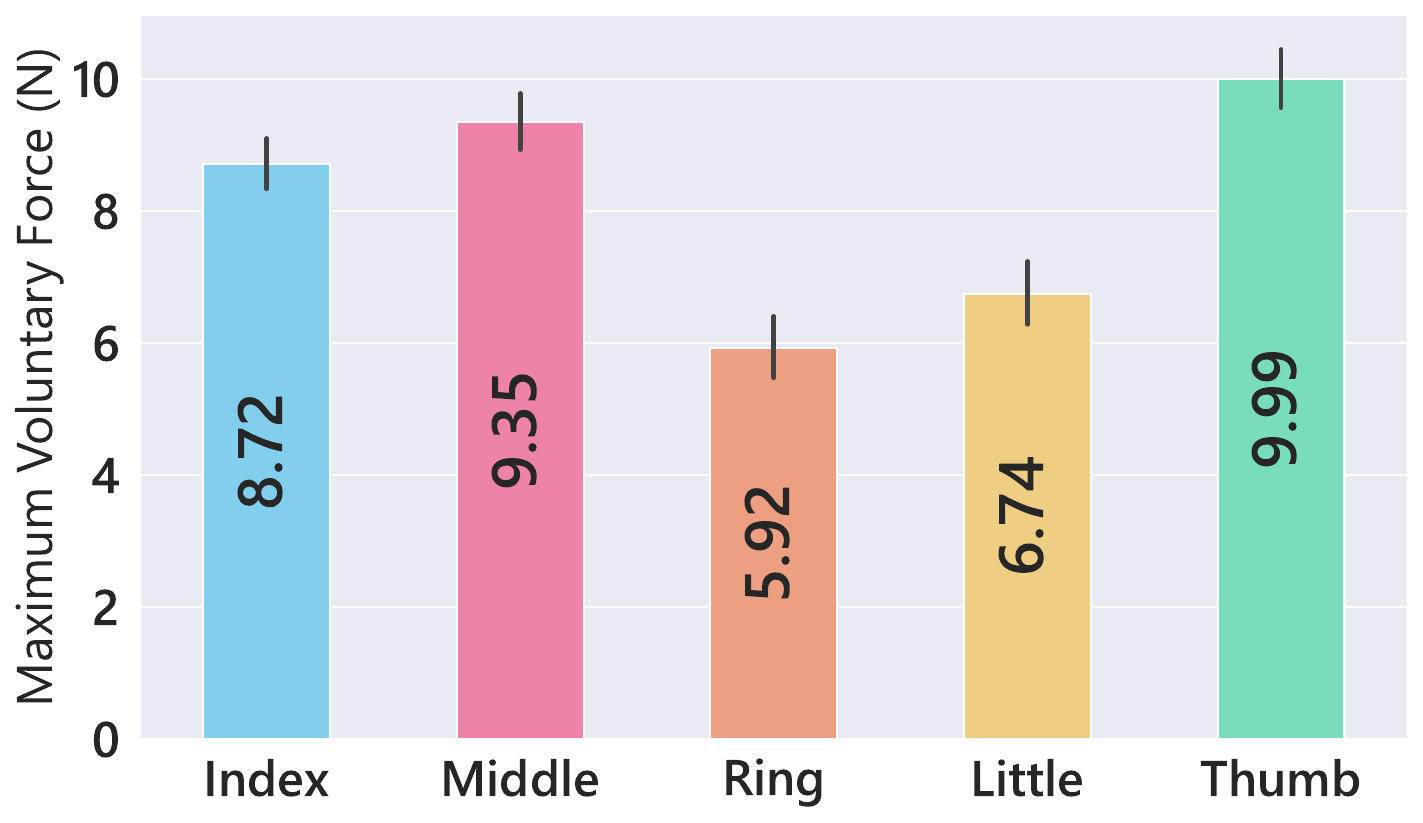}
  \captionof{figure}{Average force values (N) for all fingers across all objects and participants}
  \label{fig:force_thresholds_finger}
\end{minipage}

% \subsubsection{\textbf{Force Thresholds}}

% \setlength{\columnsep}{10pt}
% \begin{wrapfigure}{r}{0.5\textwidth}
% \vspace{-16pt}
%   \includegraphics[width=1\linewidth]{figures/pressure_thresholds_fingers_v2.pdf}
%     \caption{Average force values (N) for all fingers across all objects and participants}
%     \label{fig:force_thresholds_finger}
%     \vspace{-12pt}
% \end{wrapfigure}

% Figure~\ref{fig:force_thresholds_finger} shows average force thresholds across fingers, objects, and participants. Mean force values were: \textsc{Thumb}: 9.96N (SD = 4.14N), \textsc{Index}: 8.71N (SD = 3.58N), \textsc{Middle}: 9.20N (SD = 4.31N), \textsc{Ring}: 5.62N (SD = 4.28N), \textsc{Little}: 5.43N (SD = 4.54N). A repeated-measures ANOVA on the aligned-ranked data revealed a significant main effect of \textsc{Finger} on \textsc{Force Thresholds} ($F_{4,44} = 9.28, p < 0.0001$). Post-hoc comparisons (paired $t$-tests, Holm-corrected) showed that the fingers separated into a high-force cluster (\textsc{Thumb}, \textsc{Index}, \textsc{Middle}) and a low-force cluster (\textsc{Ring}, \textsc{Little}). Fingers within the same cluster did not differ significantly, whereas every high- versus low-cluster comparison was significant ($p < 0.05$). In particular, \textsc{Thumb} produced significantly higher force than \textsc{Ring} ($p < 0.001$) and \textsc{Little} ($p < 0.01$), but did not differ significantly from \textsc{Index} or \textsc{Middle}.
\vspace{0.15em}
\noindent cluster (\textsc{Ring}, \textsc{Little}). Fingers within the same cluster did not differ significantly, whereas every high- versus low-cluster comparison was significant ($p < 0.05$). In particular, \textsc{Thumb} produced significantly higher force than \textsc{Ring} ($p < 0.001$) and \textsc{Little} ($p < 0.01$), but did not differ significantly from \textsc{Index} or \textsc{Middle}.

To examine the effect of \textsc{Object}, we analyzed each finger across all objects (Figure~\ref{fig:force_comparisons_study_1}, top). No significant object effect was observed for any finger: \textsc{Thumb} ($F_{6,66} = 1.05, p = 0.40$), \textsc{Index} ($F_{6,66} = 1.18, p = 0.33$), \textsc{Middle} ($F_{6,66} = 0.78, p = 0.59$), \textsc{Ring} ($F_{6,66} = 0.96, p = 0.46$), \textsc{Little} ($F_{6,66} = 0.80, p = 0.57$), and the \textsc{Object} main effect was likewise non-significant. However, a significant \textsc{Finger} $\times$ \textsc{Object} interaction was found ($F_{24,264} = 5.26, p < 0.001$, Greenhouse-Geisser corrected), indicating that the magnitude of the inter-finger differences varied across objects.

% These results indicate that (i) force thresholds differ significantly across fingers, separating into a high-force group (\textsc{Thumb}, \textsc{Index}, \textsc{Middle}) and a low-force group (\textsc{Ring}, \textsc{Little}), and (ii) grasp type did not significantly change the overall force level or the force of any individual finger, although a significant \textsc{Finger} $\times$ \textsc{Object} interaction shows that the relative differences between fingers were modulated by the grasped object. Overall, these results characterize substantial differences in finger force capability while maintaining the tested grasps, with the thumb, index, and middle fingers forming a higher-force group.

\revise{These results indicate that (i) force thresholds differed significantly across fingers, separating into a high-force group (\textsc{Thumb}, \textsc{Index}, \textsc{Middle}) and a low-force group (\textsc{Ring}, \textsc{Little}), and (ii) grasp type did not significantly affect overall force levels or the force of individual fingers. However, a significant \textsc{Finger} $\times$ \textsc{Object} interaction indicates that relative differences between fingers varied with the grasped object. Overall, these findings demonstrate substantial differences in finger force capability while holding objects in different grasps.
}

% Overall, this suggests that pressure input remains reasonably robust under busy-hand conditions.

\subsubsection{\textbf{Comfort Ratings}}

\begin{figure}[!h]
    \centering
    \includegraphics[width=1\linewidth]{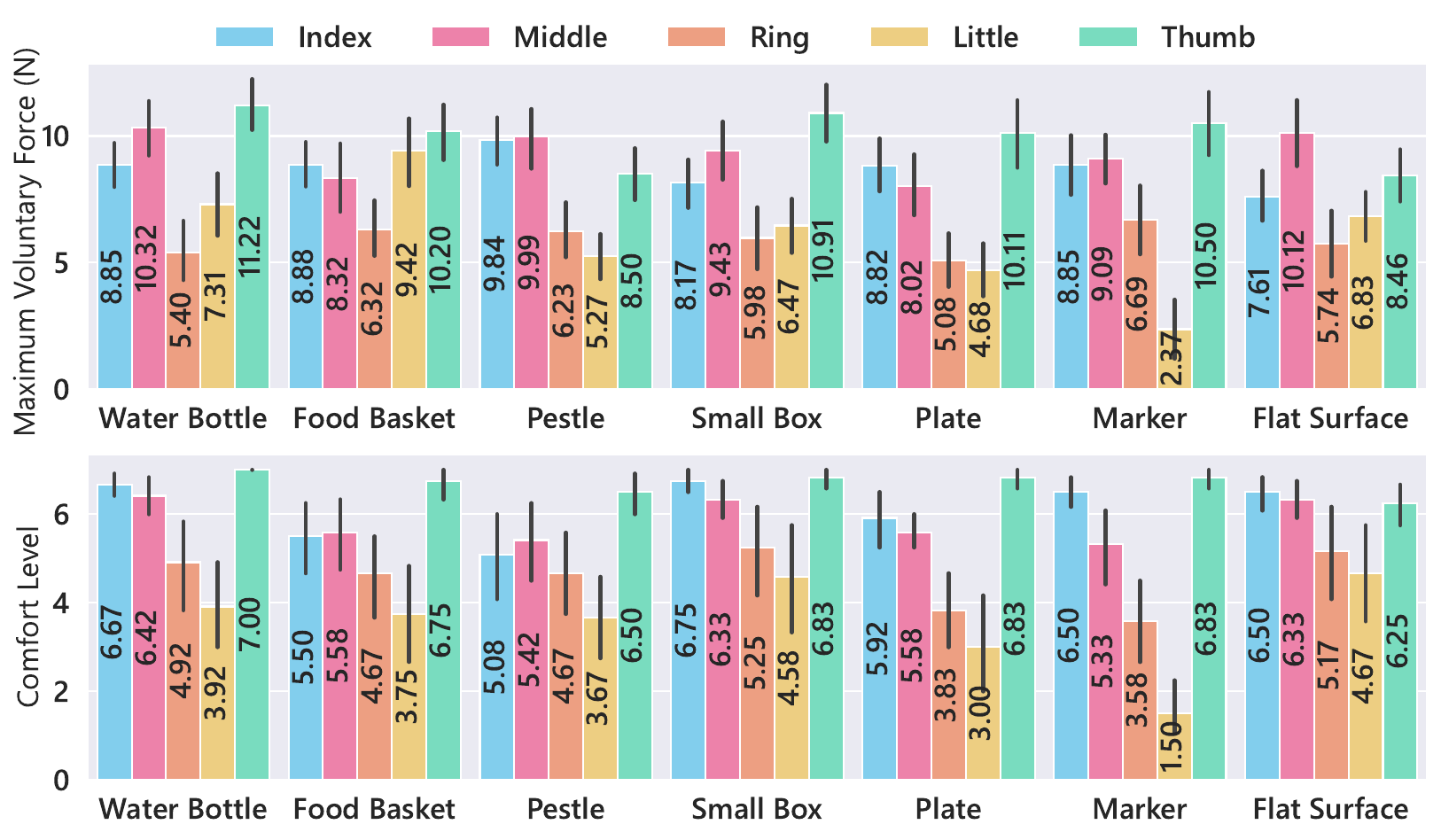}
    \caption{Average force values (top, in N) and self-reported comfort ratings (bottom, 7-point Likert scale) for each finger, averaged across objects and participants.}
    \label{fig:force_comparisons_study_1}
    \vspace{-4mm}
\end{figure}

Figure~\ref{fig:force_comparisons_study_1} (bottom) shows comfort ratings. Mean ratings were: \textsc{Thumb}: 6.71 (SD = 0.61), \textsc{Index}: 6.13 (SD = 1.21), \textsc{Middle}: 5.85 (SD = 1.19), \textsc{Ring}: 4.58 (SD = 1.81), \textsc{Little}: 3.58 (SD = 2.06). A Friedman test showed a significant effect of \textsc{Finger} on comfort ($\chi^2(4) = 217.16, p = 7.66 \times 10^{-46}$). Nemenyi post-hoc tests indicated that \textsc{Thumb} differed significantly from \textsc{Middle} ($p < 0.01$), \textsc{Ring}, and \textsc{Little} ($p < 0.0001$), but not from \textsc{Index} ($p = 0.14$). \textsc{Index} and \textsc{Middle} did not differ significantly ($p = 0.71$), while \textsc{Little} had significantly lower ratings than all others ($p < 0.0001$, except \textsc{Ring–Little}: $p = 0.03$).

%% file: experiment2_v2.tex
\section{Pressure Sensing with Busy Hands} \label{sec:busyhands}

In this section, we investigate the performance of pressure input when participants' hands are occupied while holding different types of objects. We adopt a within-subject experimental design with three independent variables: the object being held (representing different grasp types), the number of levels in the force scale, and the target force level. The following sections describe the experimental design, task, and methodology in detail.

\subsection{Task}

Following prior work, we designed a visual selection task in which participants controlled a linear on-screen cursor by modulating applied force while holding different objects. The goal was to reach a specified target level on a discrete force scale and then quickly release their finger upon selection~\cite{quick_release}.

\subsubsection{\textbf{Independent Variables}}

We manipulated the following three independent variables:

\begin{itemize}
    \item \textbf{Object (\textsc{Grasp Type})}: We used the six objects described in Section~\ref{sec:objects}, representing six distinct grasp types.
    \item \textbf{Number of levels (\textsc{Scale})}: We tested three scales composed of 8, 10, and 12 levels, in line with prior studies~\cite{pressure_widgets}.
    \item \textbf{Target force level (\textsc{Target})}: A normalized force range of 0–1000 was defined per participant, mapping minimum force to 0 and MVF to 1000. Five equidistant target levels were selected: 180 (Very Low), 360 (Low), 540 (Medium), 720 (High), and 900 (Very High)~\cite{augmenting_mouse_with_pressure_input_chi07, pressure_widgets}. The range was divided based on the selected \textsc{Scale}, with each target corresponding to a discrete level.
\end{itemize}

\subsubsection{\textbf{Dependent Variables}}
We measured three dependent variables to evaluate the performance as described in prior work~\cite{pressure_widgets}:
\begin{itemize}
    \item \textbf{\textsc{Success Rate}}: Percentage of trials where the target level was successfully reached.
    \item \textbf{\textsc{Selection Time}}: Time taken to reach the target level.
    \item \textbf{\textsc{Crossings}}: Number of entries and exits from the target level before final selection.
\end{itemize}

\subsubsection{\textbf{Choice of Finger}}
% We selected the \textsc{Thumb} as the operating finger for this experiment for three reasons. First, our prior study showed that the \textsc{Thumb} exhibits the highest force thresholds across participants. Second, comfort ratings indicated that it is the most preferred finger for pressure input in busy-hand scenarios. Third, prior work employed the thumb for pressure input in handheld contexts, such as with a mouse~\cite{augmenting_mouse_with_pressure_input_chi07} and mobile devices~\cite{pressure_input_bimanual_interaction_chi14}.

We selected the \textsc{Thumb} as the operating finger for this experiment for three reasons. First, in our prior study the \textsc{Thumb} produced the highest mean force thresholds and fell within the high-force group of fingers (together with \textsc{Index} and \textsc{Middle}), reliably exceeding the \textsc{Ring} and \textsc{Little} fingers; it was also the strongest finger for the majority of participants. Second, the \textsc{Thumb} received the highest comfort ratings overall, significantly higher than the \textsc{Middle}, \textsc{Ring}, and \textsc{Little} fingers, and comparable to the \textsc{Index}, making it among the most preferred fingers for pressure input in busy-hand scenarios. Third, prior work employed the thumb for pressure input in handheld contexts, such as with a mouse~\cite{augmenting_mouse_with_pressure_input_chi07} and mobile devices~\cite{pressure_input_bimanual_interaction_chi14}. \revise{This design does not establish that the thumb is universally superior to other fingers for pressure selection; rather, this study characterizes the performance of thumb-based pressure input. Comparative performance across fingers is left for future work.}

\noindent
\begin{minipage}[t]{0.62\textwidth}
  \vspace{0pt} % Aligns the top of the text minipage with the top of the image
  \setlength{\parindent}{1em}

  \subsection{Method}
  
  We recruited 12 participants (6 male, 6 female; mean age = 23.58, SD = 5.45). Participation was voluntary. For each object, participants held it naturally with their dominant hand. The FSR sensor was placed at the most accessible location near the \textsc{Thumb}, with placement individualized to account for variations in hand size~\cite{squish_this_chi21}.
  
  Participants first applied their MVF, with FSR readings averaged over a 50 ms window to compute a participant-specific maximum force. This value defined the upper bound of the force scale, ensuring comfortable interaction. Observed MVFs were: 13.43N $\pm$ 3.28N (cylindrical), 14.00N $\pm$ 3.22N (hook), 15.31N $\pm$ 1.66N (spherical), 14.74N $\pm$ 2.34N (palmar), 13.92N $\pm$ 3.12N (lateral), and 15.12N $\pm$ 2.20N (tip), with no systematic differences across objects.

  In each trial, participants reached the target force level indicated by a visual label, briefly stabilized to confirm their selection, and then quickly released their finger. We determined the target level selection using the algorithm proposed
  
\end{minipage}% <--- Note the percent sign here to prevent a line break
\hfill
\begin{minipage}[t]{0.35\textwidth}
  \vspace{0pt} % Aligns the top of the image minipage with the top of the text
  \centering
  \includegraphics[width=\textwidth]{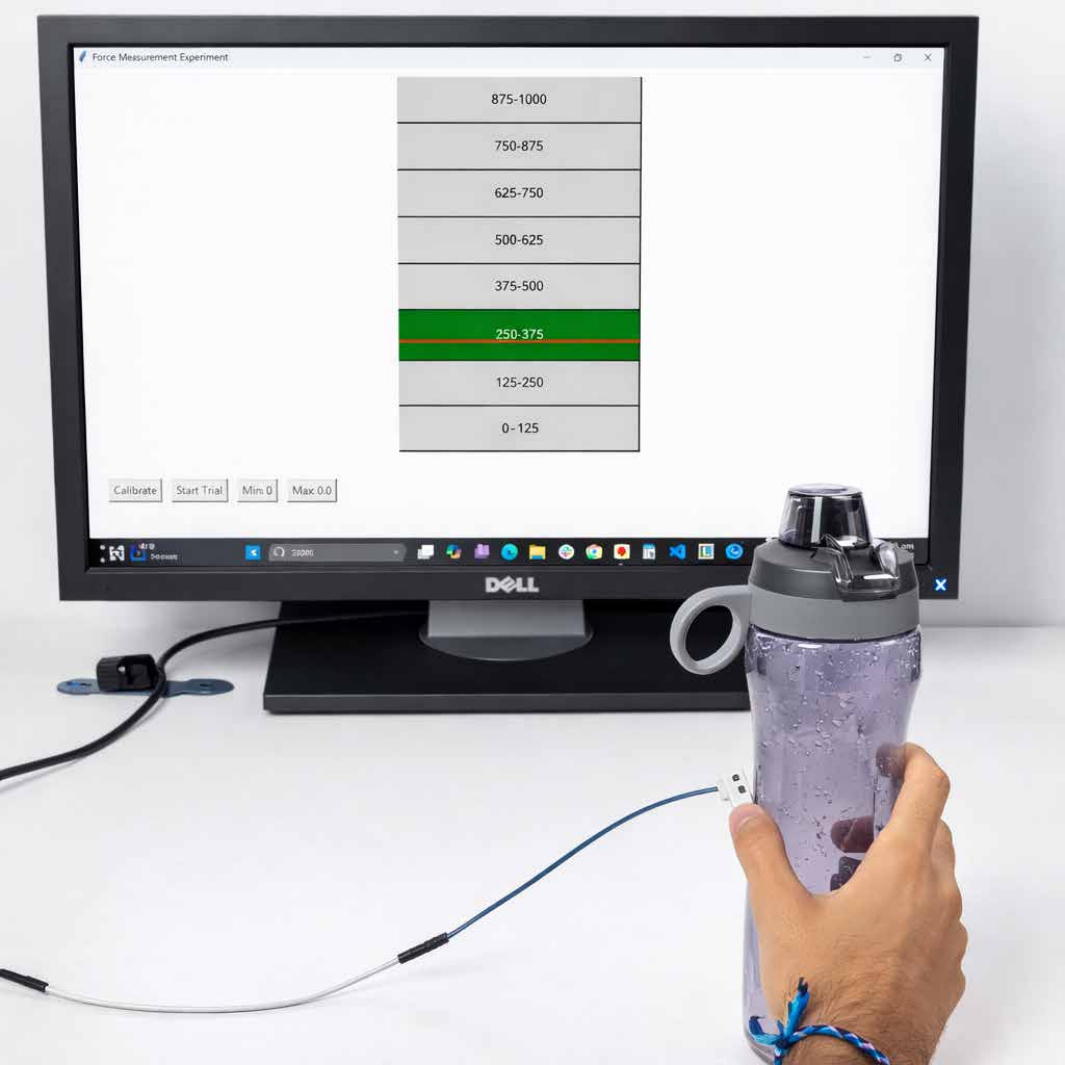}
  \captionof{figure}{Participant performing the visual pressure-selection task while holding the object with the dominant hand. The participant modulates thumb force on the FSR to move the on-screen cursor toward the indicated target level.}
  \label{fig:experimental_setup}
\end{minipage}

\vspace{0.15em}
\noindent by Corsten et al. \cite{quick_release}, which averages the sensor output over a 50 ms window ending 240 ms before the end of the movement. We counterbalanced the order of objects and scales independently: first, we counterbalanced the order of objects, and within each object, we also counterbalanced the order of scales. Prior to the main experiment, participants completed a training phase in which they practiced targeting all 5 targets in the 8-level setting with the sensor placed on a flat surface (i.e., without holding an object).

After training, participants performed the task consecutively for each object. For each \textsc{scale}, participants selected all target force levels four times in a randomized order. In total, the experiment had 12 (Participants) $\times$ 6 (Objects) $\times$ 3 (Scales) $\times$ 5 (Targets) $\times$ 4 (Repetitions) = 4320 trials. Each session lasted approximately one hour.

\revise{This study was also reviewed and approved by our institutional ethics board. All participants provided informed consent before participation and were informed that participation was voluntary and that they could withdraw at any time. Participants were given breaks throughout the experiment, and the study was stopped if a participant experienced discomfort.}

\subsection{Results}

We conducted a data-driven analysis to evaluate success rate, selection time, and crossings across all objects. To assess the effect of pressure levels, we further analyzed these metrics across five \textsc{Target} force levels: \textsc{Very Low}, \textsc{Low}, \textsc{Medium}, \textsc{High}, and \textsc{Very High}. For statistical analysis, since our data did not follow a normal distribution, we applied the Aligned Rank Transform (ART)~\cite{mansouri2004aligned,wobbrock2011aligned,art_multi-factor_uist21}.  We aligned and ranked the data with the ART tool~\cite{wobbrock2011aligned} and applied the Greenhouse-Geisser where sphericity was violated.

\subsubsection{\textbf{Success Rate}} 

\begin{figure}[!h]
    \centering \includegraphics[width=1\linewidth]{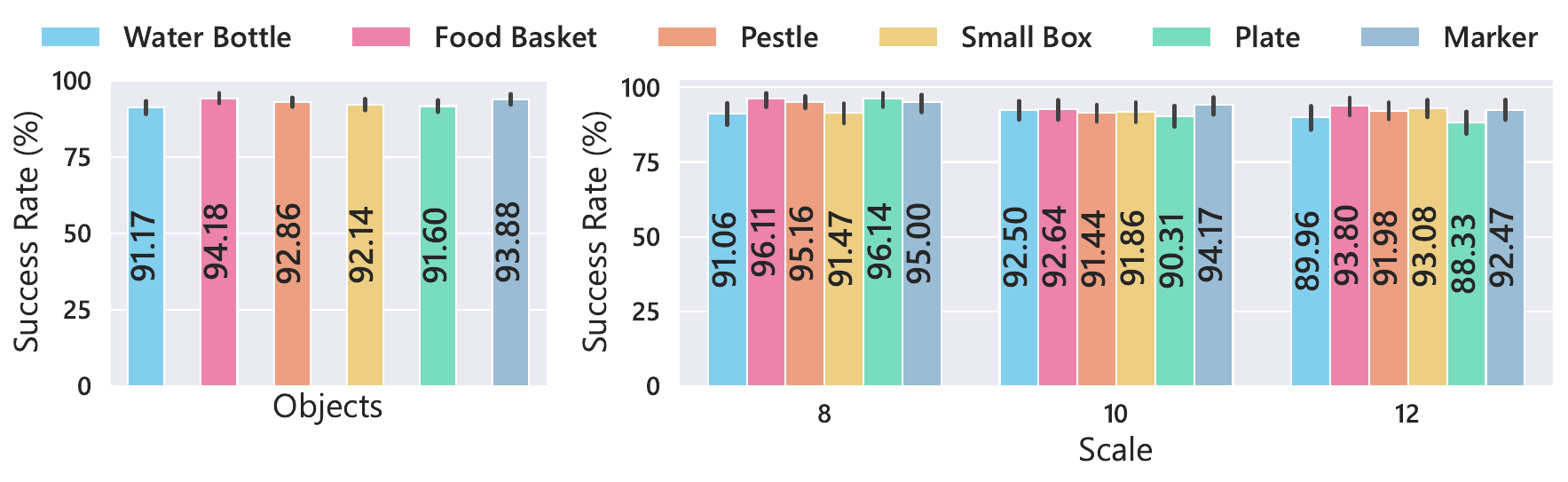}
     \vspace{-20pt}
    \caption{Average \textsc{Success Rate} across \textsc{objects} (left) and across \textsc{objects} by \textsc{scale} (right), averaged over all participants.}
    \label{fig:success_rate_study_2}
    \vspace{-12pt}
\end{figure}

Figure~\ref{fig:success_rate_study_2} shows consistently high accuracy across all objects, with an overall mean success rate of 92.7\%. Mean success rates for the \textsc{Scales} were 94.48\% (8), 92.07\% (10), and 91.31\% (12). \revise{We evaluated the effects of \textsc{Scale} (8, 10, 12) and \textsc{Object} (Food Basket, Plate, Water Bottle, Marker, Pestle, Small Box) on \textsc{Success Rate} using a two-way within-subjects Aligned Rank Transform (ART) ANOVA~\cite{wobbrock2011aligned}. As Mauchly's test indicated a violation of sphericity for \textsc{Scale} on the aligned-rank data ($\chi^2(2) = 9.38, p = 0.0092$), we applied a Greenhouse-Geisser correction ($\varepsilon = 0.62$). Our analysis revealed a significant main effect of \textsc{Scale} (${F_{1.24,13.68}=6.86,\ p=0.0161}$), with a robust partial effect size (\(\eta_p^2 = 0.38\))~\cite{cohen2013statistical}. Post-hoc power analysis confirmed that our sample size (\(N = 12\)) achieved a high statistical power of \(98.70\%\) for this effect magnitude (\(\text{Cohen's } f = 0.7900\)) at \(\alpha = 0.05\)~\cite{faul2007g}. Conversely, there was no significant main effect of \textsc{Object} (\(F_{5, 55} = 0.22,\ p = 0.9519\)), and the interaction effect between \textsc{Scale} and \textsc{Object} was entirely non-significant (\(F_{10, 110} = 1.07,\ p = 0.3933\)). This shows that while interaction success was uniformly unaffected by grasp/object type, it was significantly degraded by higher scale complexity across all evaluated conditions.}

%A Friedman test revealed a significant effect of \textsc{Scale} on \textsc{Success Rate} ($\chi^2(2) = 8.65, p = 0.01$), with a Nemenyi post-hoc test indicating a significant difference between 8 and 12 levels ($p = 0.01$). No significant differences across objects were observed within individual scales (\textsc{Scale 8}: $\chi^2(5) = 6.95, p = 0.22$; \textsc{Scale 10}: $\chi^2(5) = 1.80, p = 0.88$; \textsc{Scale 12}: $\chi^2(5) = 4.30, p = 0.51$). Overall, we did not detect a significant effect of \textsc{Object} on \textsc{Success Rate}, whereas \textsc{Scale} showed a significant effect, with the lowest scale (8) yielding significantly higher success than the highest (12), although all scales maintained mean success rates above 90\%.

% \vspace{-2mm}

\subsubsection{\textbf{Selection Time}}

\begin{figure}[!h]
    \centering \includegraphics[width=1\linewidth]{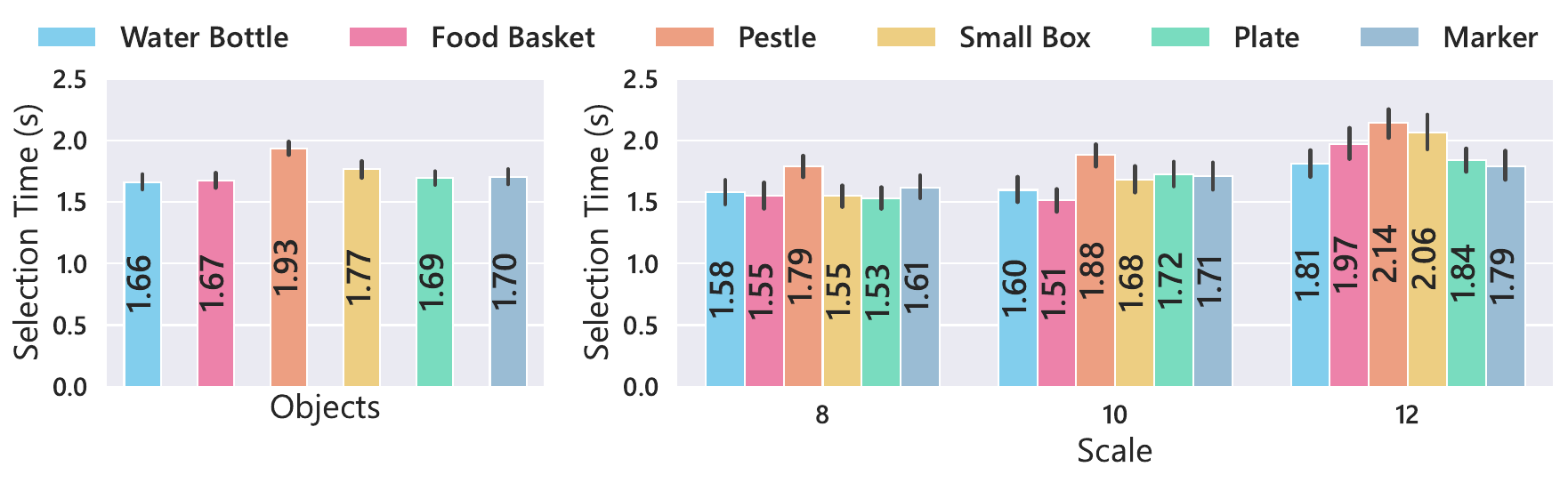}
     \vspace{-20pt}
    \caption{Average \textsc{Selection Time} across \textsc{objects} (left) and across \textsc{objects} by \textsc{scale} (right), averaged over all participants.}
    \label{fig:selection_time_study_2}
    \vspace{-12pt}
\end{figure}

Figure~\ref{fig:selection_time_study_2} (left) shows similar selection times across most objects ($\sim$1.7s), with slightly higher times for the Pestle ($\sim$1.9s) and Small Box ($\sim$1.8s). Figure~\ref{fig:selection_time_study_2} (right) shows selection times across \textsc{Scales}, with means of 1.59s (8), 1.69s (10), and 1.93s (12). \revise{An Aligned Rank Transform (ART) two-way repeated-measures ANOVA revealed a significant main effect of \textsc{Scale} (\(F_{2,22}=41.1, p<0.001, \eta_G^2=0.09, f=0.33\)), indicating that participants' performance varied across scale levels. No significant main effect of \textsc{Object} was observed (\(F_{5,55}=0.31, p=0.903, \eta_G^2=0.005, f=0.07\)), suggesting that selection times were comparable across objects when averaged across scales. A significant but comparatively small \textsc{Scale} $\times$ \textsc{Object} interaction was found (\(F_{10,110}=2.14, p=0.027, \eta_G^2=0.019, f=0.139\)), indicating that the influence of scale varied modestly across objects. To further examine the interaction, we conducted simple main-effects analyses using a Bonferroni-adjusted significance threshold of \(\alpha=0.0083\). A significant effect of \textsc{Scale} was observed only for the Food Basket (\(F_{2,22}=6.42, p=0.006, \eta_G^2=0.070, f=0.273\)). No significant effects of \textsc{Scale} were found for the Marker (\(F_{2,22}=2.67, p=0.092, \eta_G^2=0.031, f=0.178\)), Pestle (\(F_{2,22}=0.14, p=0.872, \eta_G^2=0.002, f=0.050\)), Plate (\(F_{2,22}=0.35, p=0.711, \eta_G^2=0.002, f=0.040\)), Small Box (\(F_{2,22}=0.29, p=0.749, \eta_G^2=0.003, f=0.052\)), and Water Bottle (\(F_{2,22}=1.83, p=0.184, \eta_G^2=0.018, f=0.135\)). Collectively, these findings indicate that scale exerted a robust and generally consistent influence on participants' performance across objects. Although the Scale $\times$ Object interaction was statistically significant, its effect size was small relative to the main effect of Scale, suggesting limited object-specific modulation. The \textsc{Food Basket} condition primarily drove the interaction, showing greater sensitivity to scale changes than the other objects.}

% A Friedman test revealed a significant effect of \textsc{Scale} on \textsc{Selection Time} ($\chi^2(2) = 20.17, p = 4.18 \times 10^{-5}$). Nemenyi post-hoc tests showed a significant difference only for the 8–12 \textsc{scale} pair ($scale_{8-12}: p<0.0001$; $scale_{8-10}: p=0.06$). No significant effect of \textsc{Object} was observed within individual scales (\textsc{Scale 8}: $\chi^2(5) = 2.52, p = 0.77$; \textsc{Scale 10}: $\chi^2(5) = 5.38, p = 0.37$; \textsc{Scale 12}: $\chi^2(5) = 4.05, p = 0.54$). Overall, consistent with success rate results, \textsc{Scale} significantly affected selection time, whereas we did not detect a significant effect of \textsc{Object}, with higher scales increasing selection time.

% \vspace{-2mm}
\subsubsection{\textbf{Number of Crossings}}

\begin{figure}[!h]
    \centering \includegraphics[width=1\linewidth]{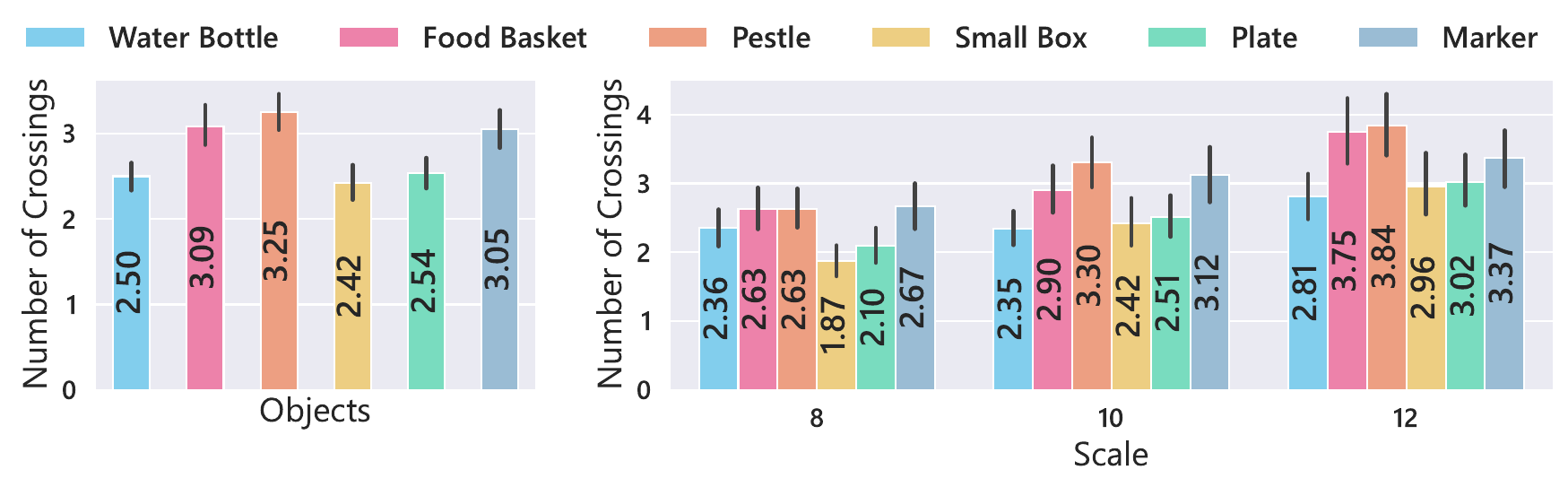}
     \vspace{-20pt}
    \caption{Average \textsc{Number of Crossings} across \textsc{objects} (left) and across \textsc{objects} by \textsc{scale} (right), averaged over all participants.}
    \label{fig:number_of_crossings_study_2}
    \vspace{-12pt}
\end{figure}

Figure \ref{fig:number_of_crossings_study_2} (left) shows the average number of crossings across different objects. Overall, it shows low crossing counts across objects, ranging from 2.4 (\textsc{Small Box}) to 3.3 (\textsc{Pestle}). Figure~\ref{fig:number_of_crossings_study_2} (right) shows that crossings increase consistently with \textsc{Scale} across most objects. Mean crossings for the \textsc{Scales} were 2.37 (8), 2.79 (10), and 3.29 (12).

\revise{An Aligned Rank Transform (ART) two-way repeated-measures ANOVA revealed a significant main effect of \textsc{Scale} ($F_{2,22}=33.73, p<0.001, \eta_G^2=0.09, f=0.33$), indicating that participants' performance varied across scale levels.} 
\revise{
No significant main effect of \textsc{Object} was observed ($F_{5,55}=1.990, p=0.09, \eta_G^2=0.05, f=0.23$), suggesting that performance were comparable across objects when averaged across scales. No significant $\textsc{Scale} \times \textsc{Object} $ interaction was found after Greenhouse-Geisser correction $(F_{10,110}=2.24, p_{GG}=0.103, \eta_G^2=0.023, f=0.152)$.}

%A Friedman test revealed a significant effect of \textsc{Scale} on \textsc{Crossings} ($\chi^2(2) = 20.67, p = 3.25 \times 10^{-5}$). Nemenyi post-hoc tests showed significant differences between 8-10 and 8-12 \textsc{scale} pairs ($scale_{8-10}: p=0.01, scale_{8-12}: p<0.0001 $). A significant effect of \textsc{Object} was found within \textsc{Scale 8} ($\chi^2(5) = 12.06, p = 0.03$), although Nemenyi post-hoc tests revealed no significant pairwise differences; no such effect was observed within \textsc{Scale 10} ($\chi^2(5) = 8.07, p = 0.15$) or \textsc{Scale 12} ($\chi^2(5) = 3.06, p = 0.69$). Overall, \textsc{Scale} significantly affected crossings, with higher levels leading to increased crossings; \textsc{Object} effects were largely absent, apart from an isolated omnibus effect within \textsc{Scale 8} that showed no significant pairwise differences.

\subsection{\textbf{Impact of \textsc{Target}}}

\subsubsection{\textbf{Success Rate}}

\begin{figure}[!h]
    \centering
    \includegraphics[width=1\linewidth]{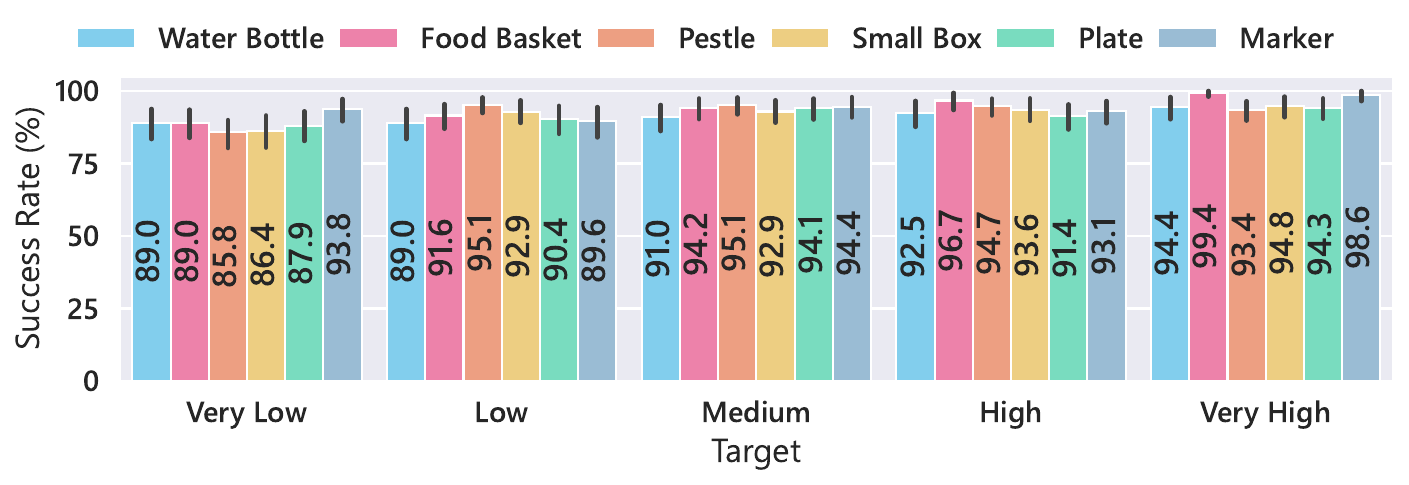}
    \vspace{-12pt}
    \caption{\textsc{Success Rate} across all \textsc{objects} and \textsc{targets} for all participants}
    \label{fig:success_rate_c_study_2}
    \vspace{-12pt}
\end{figure}

Figure~\ref{fig:success_rate_c_study_2} shows success rates across \textsc{Target} levels and objects. Overall, performance was high, ranging from $\sim$85\% at lower targets to above 95\% at higher targets. Differences across objects were small, with slightly lower success for \textsc{Pestle} and \textsc{Small Box} at \textsc{Very Low}, and near-ceiling performance for \textsc{Food Basket} and \textsc{Marker} at \textsc{Very High}. Mean success rates were 88.49\% (\textsc{Very Low}), 91.45\% (\textsc{Low}), 93.63\% (\textsc{Medium}), 93.86\% (\textsc{High}), and 95.66\% (\textsc{Very High}). \revise{An Aligned Rank Transform (ART) repeated-measures ANOVA revealed a significant main effect of \textsc{Target} (\(F_{4,44}=5.59, p=0.001\)), indicating that success rates varied across target levels. No significant main effect of \textsc{Object} was observed (\(F_{5,55}=0.50, p=0.776\)), and the \textsc{Target} \(\times\) \textsc{Object} interaction was also not significant (\(F_{20,220}=0.97, p=0.494\)). These findings suggest that target level influenced participant success, whereas object type did not. Moreover, the effect of target level on success was consistent across the tested objects.}

%A Friedman test showed a significant effect of \textsc{Target} on \textsc{Success Rate} ($\chi^2(4) = 14.35, p = 0.01$), with Nemenyi post-hoc tests indicating significant differences between \textsc{Very Low} and both \textsc{Medium} ($p = 0.04$) and \textsc{Very High} ($p = 0.01$). No significant effect of \textsc{Object} was observed within individual targets (\textsc{Very Low}: $\chi^2(5) = 5.34, p = 0.38$; \textsc{Low}: $\chi^2(5) = 2.18, p = 0.82$; \textsc{Medium}: $\chi^2(5) = 1.80, p = 0.88$; \textsc{High}: $\chi^2(5) = 2.84, p = 0.72$), except for \textsc{Very High} ($\chi^2(5) = 12.01, p = 0.03$). However, Nemenyi post-hoc tests revealed no significant differences at \textsc{Very High}, suggesting the effect does not translate to object-level differences.

Overall, higher target levels yielded more reliable performance, suggesting that assigning higher target force levels may improve robustness in pressure-based microinteractions.

\subsubsection{\textbf{Selection Time}}

\begin{figure}[!h]
    \centering
    \includegraphics[width=1\linewidth]{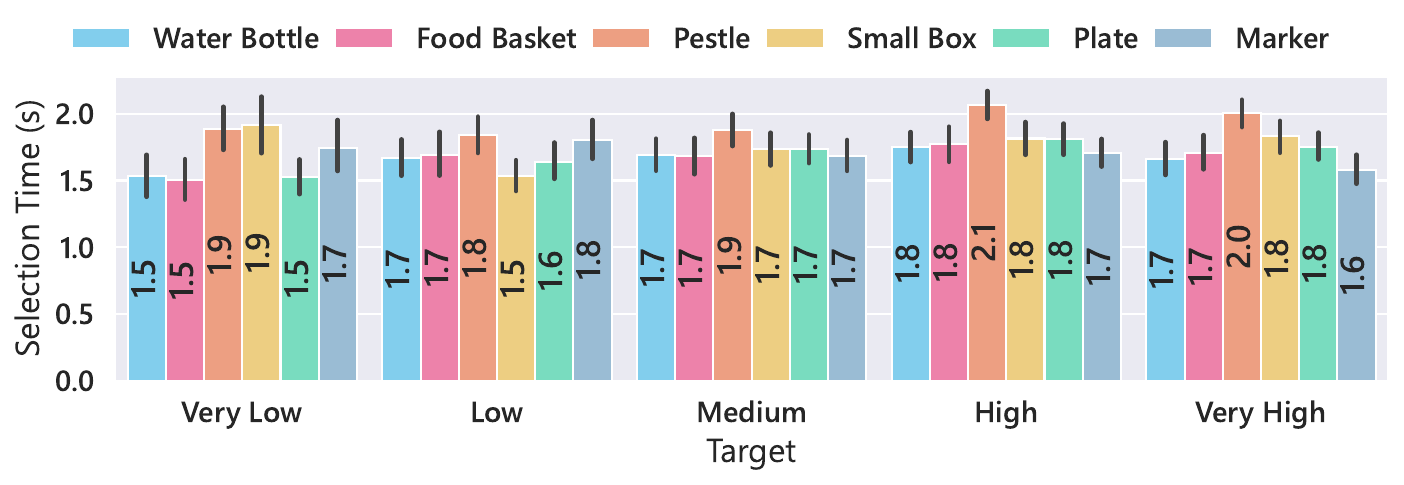}
    \vspace{-12pt}
    \caption{\textsc{Selection Time} across all \textsc{objects} and \textsc{targets} for all participants}
    \label{fig:selection_time_c_study_2}
    \vspace{-12pt}
\end{figure}

Figure~\ref{fig:selection_time_c_study_2} shows selection times across \textsc{Target} levels and objects. Times were broadly similar across objects at all target levels ($\sim$1.5–1.9s), with the \textsc{Pestle} tending to be slightly slower at higher targets (up to $\sim$2.1s for \textsc{High}). Mean selection times were 1.69s (\textsc{Very Low}), 1.69s (\textsc{Low}), 1.73s (\textsc{Medium}), 1.82s (\textsc{High}), and 1.77s (\textsc{Very High}). \revise{An Aligned Rank Transform (ART) repeated-measures ANOVA revealed no significant main effect of \textsc{Target} (\(F_{4,44}=1.13, p=0.357\)), no significant main effect of \textsc{Object} (\(F_{5,55}=0.41, p=0.838\)), and no significant \textsc{Target} $\times$ \textsc{Object} interaction (\(F_{20,220}=1.53, p=0.073\)). These results indicate that selection time remained largely stable across target levels and object types, with no statistically reliable differences between conditions.}

%A Friedman test showed no significant effect of \textsc{Target} on \textsc{Selection Time} ($\chi^2(4) = 8.73, p = 0.07$). No significant effect of \textsc{Object} was observed within any individual target (\textsc{Very Low}: $\chi^2(5) = 7.05, p = 0.22$; \textsc{Low}: $\chi^2(5) = 7.19, p = 0.21$; \textsc{Medium}: $\chi^2(5) = 1.30, p = 0.93$; \textsc{High}: $\chi^2(5) = 3.76, p = 0.58$; \textsc{Very High}: $\chi^2(5) = 8.48, p = 0.13$). Overall, \textsc{Target} did not significantly affect selection time, and object-level differences were not significant within any target.

\subsubsection{\textbf{Number of Crossings}}

\begin{figure}[!h]
    \centering
    \includegraphics[width=1\linewidth]{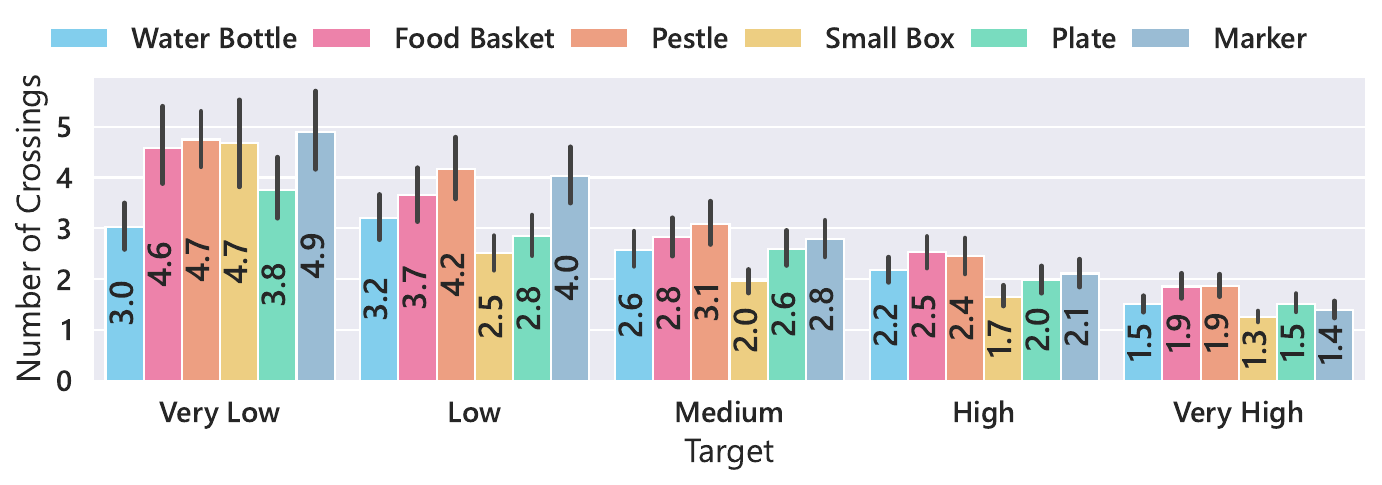}
    \vspace{-12pt}
    \caption{\textsc{Number of Crossings} across all \textsc{objects} and \textsc{targets} for all participants}
    \label{fig:number_of_crossings_b_study_2}
    \vspace{-12pt}
\end{figure}

Figure~\ref{fig:number_of_crossings_b_study_2} shows crossings across \textsc{Target} levels and objects. Crossings were highest at \textsc{Very Low} and \textsc{Low} levels ($\sim$4–5), and decreased steadily with increasing target level, reaching $\sim$1–2 at \textsc{Very High}. While this trend was consistent across objects, \textsc{Pestle} and \textsc{Food Basket} had higher crossings at lower levels, whereas \textsc{Water Bottle} and \textsc{Small Box} showed fewer.

Mean crossings were 4.29 (\textsc{Very Low}), 3.37 (\textsc{Low}), 2.66 (\textsc{Medium}), 2.19 (\textsc{High}), and 1.59 (\textsc{Very High}). \revise{An Aligned Rank Transform (ART) repeated-measures ANOVA revealed a significant main effect of \textsc{Target} (\(F_{4,44}=44.83, p<0.001, p_{GG}<0.001, \eta_G^2=0.283, f=0.628\)), indicating that crossings varied substantially across target levels. The main effect of \textsc{Object} was not significant after Greenhouse-Geisser correction (\(F_{5,55}=2.44, p=0.046, p_{GG}=0.061, \eta_G^2=0.048, f=0.225\)). Similarly, the \textsc{Target} \(\times\) \textsc{Object} interaction was not significant after correction (\(F_{20,220}=1.70, p=0.034, p_{GG}=0.160, \eta_G^2=0.036, f=0.194\)). These findings indicate that target level strongly affected crossing behaviour, whereas object type had little influence. The effect of target on crossings was largely consistent across objects.}

%A Friedman test revealed a significant effect of \textsc{Target} on \textsc{Crossings} ($\chi^2(4) = 44.77, p = 4.44 \times 10^{-9}$). Nemenyi post-hoc tests showed significant differences for the following pairs: \textsc{Very Low-Medium} ($p = 0.04$), \textsc{Very Low–High} ($p = 0.0002$), \textsc{Very Low–Very High} ($p < 0.0001$), \textsc{Low–High} ($p = 0.02$), \textsc{Low–Very High} ($p < 0.0001$), and \textsc{Medium–Very High} ($p = 0.01$). A significant effect of \textsc{Object} was found within the \textsc{Low} target ($\chi^2(5) = 13.26, p = 0.02$), although Nemenyi post-hoc tests revealed no significant pairwise differences; no significant effect was observed within the other targets (\textsc{Very Low}: $\chi^2(5) = 7.14, p = 0.21$; \textsc{Medium}: $\chi^2(5) = 5.34, p = 0.38$; \textsc{High}: $\chi^2(5) = 8.23, p = 0.14$; \textsc{Very High}: $\chi^2(5) = 8.45, p = 0.13$).
\vspace{-12pt}

\subsection{Summary of Findings}
The results across success rate, selection time, and number of crossings reveal several key patterns:

\begin{itemize}[leftmargin=*]

\item \textbf{High Accuracy Overall:} Participants achieved consistently high success rates ($\sim$92.7\% on average) across objects and grasp types, demonstrating the feasibility of pressure input with busy hands.

\item \textbf{Minimal Influence of Objects:} \textsc{Object} did not have a consistent or significant effect on performance metrics, indicating robustness of pressure input across varied geometries.

\item \textbf{Impact of Scale:} Increasing \textsc{Scale} slightly reduced success rates (94.48\% at 8 levels to 91.31\% at 12) and increased both selection time and crossings, highlighting it as a primary factor influencing performance.

\item \textbf{Effect of Target Force Levels:} Success rates improved with higher \textsc{Target} levels (88.5\% at Very Low to 95.7\% at Very High), while crossings decreased, indicating greater input stability at higher forces.

\item \textbf{Selection Time Stability:} Selection times remained relatively stable ($\sim$1.6–1.9s) across conditions, with no significant object-level differences within any target.

\item \textbf{Design Implication:} Higher target force levels are more robust and less sensitive to contextual variation, making them well-suited for microinteractions in busy-hand scenarios.

\end{itemize}

%% file: applications.tex
\section{Application Scenarios}

Our contribution is to characterize human pressure-input capabilities under busy-hand conditions, rather than to validate a specific sensing technology. \revise{While the two experiments reported earlier measured pressure using FSRs, we separately implemented three illustrative application prototypes using PressureVision++~\cite{pressurevision++}, a camera-based approach for estimating fingertip pressure from RGB images. These prototypes demonstrate that the interaction concept can be realized without instrumenting the grasped object and across different application scenarios. The quantitative force ranges and the 92.7\% success rate reported in our experiments should therefore be understood as measurements from the FSR-based studies, rather than as a direct validation of PressureVision++.}

We adopted a vision-based approach for prototyping because it offers greater scalability and flexibility: recent work has demonstrated its effectiveness for non-instrumented pressure sensing across diverse objects~\cite{pressurevision++,egotouch_uist24,Zhao_2025_CVPR}. Unlike embedded sensors, vision-based approaches eliminate the need to instrument individual objects, enable rapid prototyping across different form factors, and can leverage cameras already available in consumer devices such as HMDs, where pressure estimation can be combined with hand pose tracking~\cite{controllerPost_chi22,egotouch_uist24}. \revise{Together, the prototypes illustrate how the pressure-input capabilities characterized in our studies can be realized in practical interactive applications.}

We integrated PressureVision++ into three application scenarios:

\begin{itemize}
    \item \textbf{Controlling TV Volume While Holding a Mug:} Users adjust TV volume by modulating finger pressure on a mug while holding it naturally, enabling subtle, continuous control without additional gestures or freeing the hands.

    \item \textbf{Navigating Recipe Steps While Using a Pestle:} While grinding ingredients, users switch between recipe steps by applying pressure on the pestle handle, supporting uninterrupted, hands-on workflows without relying on touchscreens or voice input.    
    
    \item \textbf{Confirming Payment with a Credit Card:} Users confirm a digital payment by applying thumb pressure on a credit card, replacing PINs or signatures with a natural and intentional confirmation action.
\end{itemize}

\begin{figure}[!h]
    \centering
    \includegraphics[width=1\linewidth]{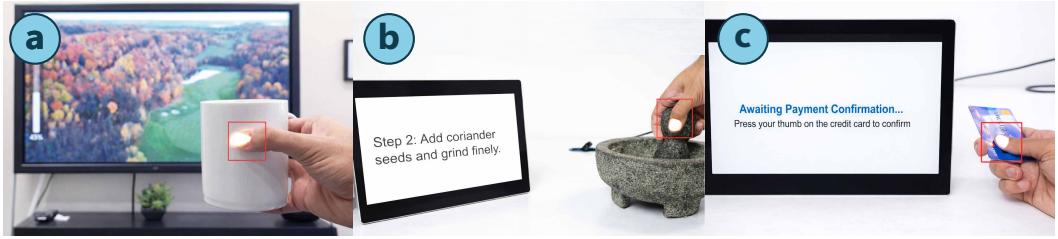}
    \caption{Application scenarios demonstrating pressure-based input while hands are occupied: (a) controlling TV volume while holding a mug, (b) navigating recipe steps while using a pestle, and (c) confirming a payment while holding the credit card.}
    \label{fig:applications}
    \vspace{-12pt}
\end{figure}

%% file: discussion.tex
\section{Discussion}

\subsection{\revise{Revisiting the Research Questions} }
\revise{Our findings answer the three research questions motivating this work. \textbf{RQ1: What force ranges can users produce while holding objects?} Participants were able to exert substantial force across all grasp conditions, although force production varied by finger and grasp type. The thumb consistently exhibited the largest MVF, followed by the middle and index fingers, suggesting a larger usable force range for interaction. }

\revise{\textbf{RQ2: Which fingers are most suitable for pressure input?} Both objective and subjective measures identified the thumb as the most suitable finger for pressure-based input. It consistently achieved the highest force output and was the most preferred and comfortable digit across grasp conditions, while the middle and index fingers also emerged as viable alternatives in some contexts.} 

\revise{\textbf{RQ3: How well can users perform pressure input with busy hands?} The results from Study 2 reveal that interaction performance was primarily constrained by force-level granularity rather than object geometry. Statistical analyses showed significant effects of \textsc{Scale} across multiple performance measures, while \textsc{object} type generally had limited influence on success rates and selection times. This suggests that users can adapt pressure input across diverse grasp configurations, provided that the number of force levels remains within a manageable range.}

\subsection{\revise{Design Implications}}
\revise{Taken together, our findings indicate that pressure can serve as a practical complementary modality for busy-hand interaction and suggest three actionable design implications.}

\begin{itemize}[leftmargin=*]
    \item \revise{\textbf{Finger-Aware Interaction Design}: Designers should carefully consider which finger is assigned to pressure-based input. The \textsc{Thumb}, which consistently combined the largest force range with the highest user preference, appears best suited for continuous pressure control and multi-level input. The \textsc{Middle} and \textsc{Index} fingers provide viable alternatives when thumb input is unavailable or engaged in grasp stabilization. In contrast, the lower force capabilities of the \textsc{Ring} and \textsc{Little} fingers may make them better suited for binary commands, mode switching, or low-granularity input. More broadly, these differences suggest opportunities for finger-specific pressure mappings that assign interaction roles according to the force capabilities and grasp functions of individual digits, potentially enabling richer pressure-based vocabularies and multi-finger interactions.}

    \item \revise{\textbf{Granularity Choice}: Designers should carefully select pressure-level granularity, as interaction performance was affected far more by \textsc{Scale} than by object geometry. In practice, increasing the number of force levels beyond moderate granularities introduces measurable performance costs, suggesting that pressure mappings should favour fewer, well-separated levels unless fine-grained control is essential.}

    \item \revise{\textbf{Transferability}: The relatively small effect of object type on performance suggests that pressure-based interactions can transfer across a wide range of everyday objects and grasp configurations. Rather than developing object-specific pressure techniques, designers may instead focus on calibration, feedback, and scale design, which appear to have a greater impact on usability.}
\end{itemize}

  Collectively, these findings suggest that robust pressure-based interfaces should be finger-aware, scale-conscious, and largely agnostic to the specific object being held.

\subsection{Microgestures and Pressure Input}

Ashbrook defines microinteractions as interactions completed in under four seconds~\cite{ashbrook2010enabling}. In busy-hand contexts, our results show that pressure input meets this criterion, with an average selection time of 1.74 s. Rather than replacing modalities such as microgestures, we position pressure input as a complementary interaction channel. This characterization is based on the distinct interaction affordances of the two modalities rather than a direct empirical comparison under matched busy-hand conditions. While microgestures support expressive, discrete commands, pressure input enables subtle, continuous, fine-grained control. Combining them could enable hybrid techniques that adapt to task demands, context, and user intent, offering a promising direction for future work.

\subsection{Finger Co-Activation and False Activations}

A common challenge in microgestures and mid-air finger interaction is co-activation, where neighbouring fingers move involuntarily during intentional gestures~\cite{schieber1991individuated}. This can cause false activations in IMU-based systems~\cite{sparseimu_tochi23,sharma2021solofinger}. Pressure-based input offers advantages here, as it is more resilient to co-activation, consistent with our findings. Differences in maximum voluntary force (MVF) across fingers can also support threshold-based filtering of unintended input. For example, higher force levels can be reliably produced only by stronger fingers (e.g., \textsc{Thumb} and \textsc{Middle}), reducing accidental activation. If sensing is contact-based, the sensor placement is also inherently finger-specific (e.g., a sensor positioned for the \textsc{Thumb} cannot be triggered by the \textsc{Middle}), further mitigating false activations.

\section{Limitations and Future Work}

\subsection{Sensor Placement and In-The-Wild Studies }
In our experiments, we positioned the force-sensitive resistors (FSRs) according to individual preference, since we used only a single sensor at a time. While effective for controlled studies, this approach does not scale well across users and object geometries. A more practical solution is to employ a pressure-sensing matrix that covers a larger contact area, enabling dynamic adaptation to different users and grasp configurations while improving robustness and generalizability. Additionally, we conducted our experiments in a controlled setting to isolate the effects of object type and grasp. Future work should evaluate pressure input in the wild during everyday tasks to assess ecological validity and practical deployment.

\subsection{Weight and Size}

We did not explicitly examine how object weight and size affect pressure-based input. However, our observations and prior work~\cite{grasping_microgestures_chi19} suggest that object size matters more than weight in shaping how users apply and regulate pressure. While our study focused on small, handheld objects typical of mobile scenarios, future work should explore larger and more varied geometries to assess the impact on precision and comfort. We also observed that for objects such as the pestle, where force is inherent to the primary task, false activations were more common at lower force levels. Despite this, success rates remained high (85.58\% at very low and 94.6\% at higher levels), suggesting that medium-to-high force thresholds are preferable for reliable input. These findings further highlight the importance of considering object–task dynamics for designing pressure microinteractions.

\subsection{\revise{Force Equilibrium and Multi-Finger Coordination}} 
\revise{ Pressure-based input while grasping an object does not occur in isolation. During object manipulation, force generated by one digit is often accompanied by compensatory forces from other digits to maintain grasp stability and satisfy force-equilibrium constraints~\cite{zatsiorsky2008multifinger,hermsdrfer2009analysis}. Because our study only measured the target finger, we cannot determine how non-target digits contributed to force regulation during pressure input. Consequently, our findings should be interpreted as characterizing the pressure-input capabilities of individual digits rather than providing a complete account of force coordination during grasping. Future work should employ multi-finger force sensing to quantify inter-finger force coupling and investigate how force-sharing strategies influence pressure-based interaction performance. }

\revise{\subsection{Grasp Constraints and Hand Pose}
Although our study examined pressure input across six representative grasp types, we did not systematically manipulate hand posture, wrist orientation, or finger placement within each grasp category. Prior biomechanics research shows that force production capabilities can vary with joint configuration, posture, contact geometry, and muscle recruitment patterns~\cite{hermsdrfer2009analysis,RADWIN01031992}. Consequently, the measured maximum voluntary forces should not be interpreted as posture-independent limits that generalize across all ways of performing a given grasp. Our goal was not to exhaustively model all biomechanical determinants of force generation, but rather to characterize pressure-input capabilities under representative busy-hand interaction scenarios. Future work should systematically vary hand and wrist posture, object geometry, and grasp configuration to better understand how these factors influence the usable force range and performance of pressure-based interaction.
}

\revise{\subsection{Vision-Based Sensing Considerations}
PressureVision++ may introduce additional sources of variability, including estimation error, image quality, occlusion, hand/object configuration, and differences in effective sensing resolution. Its ability to support the finer force-scale conditions used in our FSR experiments was not evaluated in this work. In our prototype implementations, we found that coarser discretizations (e.g., 3–4 levels) were practical for demonstrating force-based interactions; however, this observation is implementation experience rather than a controlled evaluation. Future work should compare sensing accuracy, resolution, and interaction performance across FSR and vision-based sensing.}

%% file: conclusion.tex
\section{Conclusion}

In this work, we systematically investigated the feasibility of pressure-based input in busy-hand scenarios where users interact while holding everyday objects. Across two user studies spanning six fundamental grasp types, we characterized finger-specific maximum voluntary forces, examined user preferences, and evaluated interaction performance through success rates, selection times, and number of crossings. Our results show that users can reliably perform pressure input with busy hands, achieving a high success rate (92.7\%) and low average selection time (1.74 s), even with multiple pressure levels. We further demonstrated the practical potential of this approach through three application scenarios. Taken together, these findings suggest that pressure-based input can serve as a complementary microinteraction modality alongside techniques such as microgestures, enabling efficient and unobtrusive interaction under grasp-constrained conditions.